\documentclass[letterpaper, 10pt, journal, final]{IEEEtran}

\IEEEoverridecommandlockouts                              % This command is only needed if 
\usepackage{graphicx} % for pdf, bitmapped graphics files
\usepackage{epsfig} % for postscript graphics files
\usepackage{mathptmx} % assumes new font selection scheme installed
\usepackage{times} % assumes new font selection scheme installed
\usepackage{amsmath} % assumes amsmath package installed
\usepackage{amssymb}  % assumes amsmath package installed
\usepackage{bbold} % added by ivine
\usepackage{float} % added by ivine

\title{\LARGE \bf
Inference of the Selective Auditory Attention using Sequential LMMSE Estimation
}

\author{Ivine Kuruvila$^{1}$, Kubilay Can Demir$^{1}$, Eghart Fischer$^{2}$, Ulrich Hoppe$^{1}$% <-this % stops a space
\thanks{This work was carried out at the ENT clinic, Friedrich-Alexander-Universit\"at Erlangen-N\"urnberg (FAU), Erlangen, Germany and had received a grant from WS Audiology.}
\thanks{$^{1}$Ivine Kuruvila, Kubilay Can Demir and Ulrich Hoppe are with Department of Audiology, ENT-Clinic, Friedrich-Alexander-Universit\"at Erlangen-N\"urnberg (FAU), Erlangen, Germany.
        {\tt\small Ulrich.Hoppe@uk-erlangen.de}}%
\thanks{$^{2}$Eghart Fischer is with WS Audiology, Erlangen, Germany.}%
}

\begin{document}
\maketitle
\thispagestyle{empty}
\pagestyle{empty}

%%%%%%%%%%%%%%%%%%%%%%%%%%%%%%%%%%%%%%%%%%%%%%%%%%%%%%%%%%%%%%%%%%%%
\begin{abstract}
Attentive listening in a multispeaker environment such as a cocktail party requires suppression of the interfering speakers and the noise around. People with normal hearing perform remarkably well in such situations. Analysis of the cortical signals using electroencephalography (EEG) has revealed that the EEG signals track the envelope of the attended speech stronger than that of the interfering speech. This has enabled the development of algorithms that can decode the selective attention of a listener in controlled experimental settings. However, often these algorithms require longer trial duration and computationally expensive calibration to obtain a reliable inference of attention. In this paper, we present a novel framework to decode the attention of a listener within trial durations of the order of two seconds. It comprises of three modules: 1) Dynamic estimation of the temporal response functions (TRF) in every trial using a sequential linear minimum mean squared error (LMMSE) estimator, 2) Extract the N1$-$P2 peak of the estimated TRF that serves as a marker related to the attentional state and 3) Obtain a probabilistic measure of the attentional state using a support vector machine followed by a logistic regression. The efficacy of the proposed decoding framework was evaluated using EEG data collected from 27 subjects. The total number of electrodes required to infer the attention was four: One for the signal estimation, one for the noise estimation and the other two being the reference and the ground electrodes. Our results make further progress towards the realization of neuro-steered hearing aids.
\newline
\end{abstract}

%%%%%%%%%%%%%%%%%%%%%%%%%%%%%%%%%%%%%%%%%%%%%%%%%%%%%%%%%%%%%%%%%%%%%%%%%%%%%%%%
\section{Introduction}
The cocktail party effect or the ability of humans to focus the attention on a certain speaker among multiple speakers is fundamental to interpersonal communication. In the last decades, scientific community has actively investigated how the human brain is able to maintain uninterrupted attention in the presence of noise and interference from different speakers. Although the process by which the brain segregates multiple speakers still remains largely unknown, progresses are continuously being made by analyzing the cortical signals measured using electrocorticography (ECoG), magnetoencephalography (MEG) and electroencephalography (EEG). In recent years, EEG analyses have become the forefront of auditory attention research, thanks to its non-invasive nature, minimal effort and high temporal resolution. Several methods have been proposed that model the relationship between  multiple speakers present in an auditory scene and the elicited EEG signal \cite{bib:power2012} \cite{bib:sullivan2014} \cite{bib:fiedler2019} \cite{bib:das2020}. 

The underlying assumption in these methods is that the cortical signal track the acoustic envelope of the attended speaker stronger than that of the unattended speaker \cite{bib:sullivan2014} \cite{bib:aiken2008}. While the path between the outer ear and the EEG/MEG sensor is non-linear, linear system analysis is often used to analyse the processing of speech envelope through the auditory path. The system response function that models this pathway is termed as temporal response function (TRF) \cite{bib:lalor2010} \cite{bib:ding2012b}. Since this mapping is from speech to EEG/MEG sensors in the forward direction, TRFs fall under the category of forward models. TRFs are not limited to speech envelopes but can be used to map a linear relationship between speech spectrograms and the cortical signals, or between phonemes and the cortical signals \cite{bib:ding2012b} \cite{bib:diLiberto2015}. In EEG modality, TRFs have high temporal resolution with peaks around 100 ms and 200 ms that modulate the attentional effect \cite{bib:fiedler2019}. On the other hand, due to the assumption of linearity, EEG signals can be mapped on to the speech envelope in backward direction (backward model). The current state-of-the-art system identification method for both forward and backward model is based on least-squares (LS) \cite{bib:mesgarani2009} and the identified system can be used to deduce the attention of a listener \cite{bib:sullivan2014} \cite{bib:mirkovic2015} \cite{bib:biesmans2017} \cite{bib:fuglsang2017} \cite{bib:fiedler2017} \cite{bib:wong2018}. However, the system response function calculated using the backward model cannot be physiologically interpreted due to its implicit noise suppression \cite{bib:haufe2014}. Therefore, TRF based on the forward model using EEG signals is the focus of this paper.

Although the aforementioned studies have proven successful in estimating the selective attention reliably, they suffer from two major limitations. First, the temporal resolution required to obtain a reliable auditory attention decoding (AAD) accuracy is of the order of tens of seconds whereas a listener can switch the attention at much shorter time scales. Second, attention decoding accuracies deteriorate as the number of EEG electrodes used for analysis reduces. A recent study based on the state space model has addressed the first limitation whereby the temporal resolution was reduced to less than one second \cite{bib:miran2018}. Similarly, the effect of reducing the number of EEG electrodes on the attention decoding accuracy was investigated in \cite{bib:mirkovic2015} \cite{bib:fuglsang2017} \cite{bib:narayanan2019}.

%Second, system identification methods that rely on LS regression algorithms often suffer from poorly determined correlation matrices that are numerically unstable under inversion. The correlation matrix can be made invertible by imposing a penalty on the L1 norm (LASSO) or on the L2 norm (rigde regression) of the regression coefficients \cite{bib:hastie2001} \cite{bib:wong2018}. However, to calculate an optimal penalization/regularization parameter ($\lambda$), training on large dataset with repeated calibration is necessary. Moreover, $\lambda$ has an influence on the temporal resolution of the attention decoding. For example, a $\lambda$ that is optimal for the first 500 ms may not be optimal for the last 500 ms in a trial (a block of EEG signal) of 10 seconds duration. And oftentimes, $\lambda$ is fixed for a given subject and this fixed $\lambda$ is used for the entire analysis which may lead to sub-optimal results. These limitations hinder the application of attention decoding algorithms that require near real time processing such as brain-computer interface (BCI) or hearing aids. A recent study based on the state space model has addressed the first limitation whereby the temporal resolution was reduced to less than one second \cite{bib:miran2018}.

Here, we present a novel framework to infer the selective auditory attention of a listener. The proposed framework consists of three modules. The first module relates to the dynamic estimation of the TRF corresponding to the attended speaker (attended TRF) and the unattended speaker (unattended TRF). It is based on sequential linear minimum mean squared error (LMMSE) estimator which is an improvement of the algorithm proposed in \cite{bib:kuruvila2020}. LMMSE algorithms are based on explicitly calculating the covariance of the signal component and subsequently applying Bayesian estimation theory. In the second module, a marker related to the attention of the listener is extracted. Correlation based marker is a commonly used attention marker where the Pearson correlation coefficients between the original signals and the reconstructed signal are compared \cite{bib:fiedler2017} \cite{bib:miran2018} \cite{bib:alickovic2019}. However, it has also been suggested that the amplitude peaks of the attended TRF and the unattended TRF differ around 100 ms (N1\textsubscript{TRF}) and 200 ms (P2\textsubscript{TRF}) latencies and these peaks correlate with selective attention \cite{bib:fiedler2019} \cite{bib:ding2012b} \cite{bib:akram2016a}. As the sequential LMMSE estimator is capable of generating TRFs with high fidelity from short duration trials, we define the attention marker as the magnitude of the difference between N1\textsubscript{TRF} and P2\textsubscript{TRF} known as the N1\textsubscript{TRF}$-$P2\textsubscript{TRF} peak. In the third module, we train a support vector machine (SVM) using attention markers corresponding to the two speakers as features in order to generate a decision boundary and classify attention. Finally, a logistic regression is applied to the SVM's output to obtain a probabilistic confidence of the attentional inference.

The rest of the paper is organized as follows. In section \ref{sec:MandM}, experimental details, information about the speech stimuli and the EEG data collection procedure are provided. In section \ref{sec:AAD_framework}, the proposed attention decoding framework is developed and explained in detail. Classification results are presented in section \ref{sec:Results} and the paper is concluded with a discussion on the algorithm and the results in section \ref{sec:Discussion}.

\section{Materials and Methods}
\label{sec:MandM}

\subsection{Participants}
Twenty seven subjects who were all native German speakers took part in the study. Of these, eighteen subjects (M$_{age}$ = 58 years, range = 24 - 73 years, nine females, nine males) were hearing impaired and the remaining nine subjects  (M$_{age}$ = 32 years, range = 20 - 52 years, five females, four males) were of normal hearing. Individuals with pure tone audiometric thresholds better than 25 dB HL at octave frequencies from 125 Hz to 8 kHz were considered as normal hearing whereas individuals with pure tone audiometric thresholds worse than 25 dB HL were considered as hearing-impaired. All participants gave their written consent before the start of the experiments and the study was approved by the ethics committee at the university of Erlangen-Nuremberg. 

\subsection{Experimental Design}
\label{sec:ED}
To emulate a cocktail party effect, two streams of news spoken by two different male speakers were presented to the subject simultaneously. They were presented through two loudspeakers which were connected to a computer through Fireface UC soundcard (RME Haimhausen Germany). Loudspeakers were placed one meter away from the subject at an azimuth of +45$^\circ$ and -45$^\circ$ on the left and on the right, respectively. The total duration of the speech stimuli was 30 minutes and consisted of six segments with each segment being approximately five minutes long. Subjects were asked to focus their attention either to the left speaker or to the right speaker in the first four segments and to the other speaker in the remaining two segments. The initial attention to the left or to the right was chosen randomly across the subjects. After every five-minute segment, subjects were given a multiple choice questionnaire related to contents of the attended stream to motivate them to comply with the task.

Before the start of EEG measurements, loudness of the loudspeakers was calibrated to the most comfortable level per individual subject. This was determined by increasing the loudness of a particular loudspeaker until the subject indicates that the contents are clearly understandable. Subsequently, loudness of the other loudspeaker was adjusted to the same level. Hearing-impaired subjects were requested to remove their hearing aids before the experiment so as to avoid any non-uniform speech processing introduced by the different hearing aid manufacturers. Although participants in our experiment consisted of both hearing impaired and normal hearing subjects, it is not within the scope of this study to present a comparative analysis of both the groups. Comprehensive analyses of the effect of hearing loss on cortical tracking during selective auditory attention are provided in \cite{bib:presacco2019} \cite{bib:decruy2020} \cite{bib:fuglsang2020}.

\subsection{Speech Stimulus and EEG Recordings}
Speech stimuli used in our selective auditory experiment were taken from the slowly spoken news section of the German news website \textit{www.dw.de} (Deutsche Welle). To reduce subjects' prior knowledge, news contents were taken from the 2015-16 archive of the website. Each news was around 60 seconds long and were presented only once to the subject. The sampling rate of the speech signal was 44.1 kHz. EEG signal was recorded with the help of Brainamp MR amplifier (Brainproducts Munich, Germany) using 21 AgCl electrodes placed according to 10-20 EEG format. The sampling rate of the EEG signal was 2.5 kHz and the reference electrode was placed at the right mastoid and the ground electrode was placed on the left earlobe. As far as possible, the impedance of the electrodes were maintained under 10 k$\Omega$. The speech stimuli and the EEG signals were synchronized using a trigger signal that was send through the soundcard to the EEG amplifier.

\subsection{Data Analysis}
The speech envelope was obtained by taking the absolute value of Hilbert transform of the speech signal. As many studies on attention decoding have shown that the low frequency cortical signal under 10 Hz track the low frequency speech envelope, both the EEG and the speech envelopes were downsampled to 64 Hz \cite{bib:power2012} \cite{bib:sullivan2014}. Afterwards, they were bandpass filtered between 1-9 Hz. In time domain, the trial duration considered for analysis was 2 sec throughout this paper unless otherwise stated. Two seconds trial duration was chosen because the human brain can process of the order of two seconds of independent auditory information in working memory \cite{bib:baddeley1975}. EEG signals measured at electrodes whose impedance was larger than 10 k$\Omega$ were discarded from further investigation. All analysis were performed using MATLAB, version R2017b (The Mathworks Inc. Natick, Massachusetts).

\section{Attention Decoding Framework}
\label{sec:AAD_framework}
Our attention decoding framework consists of three modules. In the first module (section \ref{sec:TRF_Est}), response of the auditory system that generated the cortical signals when attended to a certain speaker is estimated. In the second module (section \ref{sec:EAM}), a marker corresponding to the dynamic attentional state of the listener is extracted. In the third and final module (section \ref{sec:SVM}), attention markers are fed as input to an SVM classifier to infer the current attention state. A logistic regression is then trained on the classifier output to generate soft decisions corresponding to the probability of attention to a certain speaker. 

\subsection{TRF Estimation}
\label{sec:TRF_Est}
\subsubsection{LMMSE}
TRF is the linear system response function that characterizes the auditory pathway inside the brain when a person listens to a certain speaker. It has now been established that the cortical signals track the envelope of speech during attentive listening \cite{bib:aiken2008}.

Let $\theta$, a $p\times 1$ vector, denote the linear system response function and $s$, an $N \times 1$ vector ($ N>p $), denote the input speech signal envelope. Let $r$ denote the cortical signal measured at an EEG electrode when a system with $\theta$ as TRF is stimulated by the input $s$.  Then $r$ can be expressed as 

\begin{equation} \label{eq1}
\begin{split}
r &= s*\theta + w\\
&= S\theta + w,
\end{split}
\end{equation}

where $S$ is a known matrix of order $N\times p$ which is the causal time-lagged version of input speech envelope, $w$ is an $N \times 1$ vector that captures the observation noise and the model inaccuracies and $*$ denotes the linear convolution operator. The most commonly used algorithm to estimate the system response function in \eqref{eq1} is based on LS estimation where it is assumed that $\theta$ is a deterministic variable \cite{bib:crosse2016}. As it may not always be guaranteed as deterministic, we assume $\theta$ to be a random variable with a prior probability density function (PDF) $\mathcal{N}(\mu_{\theta},C_{\theta \theta})$. Noise vector $w$ is also assumed to be a random variable with a prior PDF  $\mathcal{N}(0,C_{w})$ and independent of $\theta$. This form is known as Bayesian linear model and the estimator that minimizes the mean squared error is given as \cite{bib:mendel1995} \cite{bib:kay1997}

\begin{equation} \label{eq2}
\hat{\theta} = E(\theta|r).
\end{equation}
The posterior mean in \eqref{eq2} can be expressed as \cite{bib:kuruvila2020}

\begin{equation} \label{eq3}
\hat{\theta}  = E(\theta) + C_{\theta r} C_{rr}^{-1}(r-E(r)) 
\end{equation}
where $E(\cdot)$ corresponds the sample mean, $C_{rr}$ corresponds to the autocovariance of the observation signal and $C_{\theta r}$ corresponds to the cross-covariance between the observation and the system response. Equation \eqref{eq3} can be further expanded as \cite{bib:kuruvila2020}

\begin{equation} \label{eq4}
\hat{\theta}  = \mu_{\theta} + C_{\theta \theta}S^T(SC_{\theta \theta}S^T + C_w)^{-1}(r-S\mu_{\theta})
\end{equation}
which gives the LMMSE estimate of $\theta$. When data from multiple electrodes need to be analyzed, \eqref{eq4} must be solved separately for signals at each electrode.\\

\subsubsection{Sequential LMMSE}
LMMSE estimation or LS estimation methods assume that $\hat{\theta}$ estimated in the previous trial does not have any influence on the $\hat{\theta}$ estimated in the current trial. But as auditory attention is a continuous process, attention during the previous instance may have an influence on the attention at the current instance. In such scenarios, $\hat{\theta}$ at n\textsuperscript{th} trial can be estimated sequentially as \cite{bib:kay1997}\\

\textbf{Estimator Update:}
\begin{equation} \label{eq5}
\hat{\theta}[n] = \hat{\theta}[n-1] + K[n](r-S[n]\hat{\theta}[n-1]),
\end{equation}
where

\begin{equation} \label{eq6}
K[n] = M[n-1]S^T[n](C_w + S[n]M[n-1]S^T[n])^{-1}
\end{equation}
\textbf{Minimum MSE Matrix Update:}

\begin{equation} \label{eq7}
M[n] = (I - K[n]S[n])M(n-1)
\end{equation}

The gain factor $K[n]$ is a $p\times 1$ vector and mean squared error (MSE) $M[n]$ is a $p\times p$ matrix that corresponds to the covariance of the estimate $\hat{\theta}$ at n\textsuperscript{th} trial. A procedure to calculate the initialization parameter $\hat{\theta}[-1]$ and $M[-1]$  will be discussed in section \ref{sec:EMP}.\\

\begin{figure}[h]
\centering
\includegraphics[width=.4\textwidth]{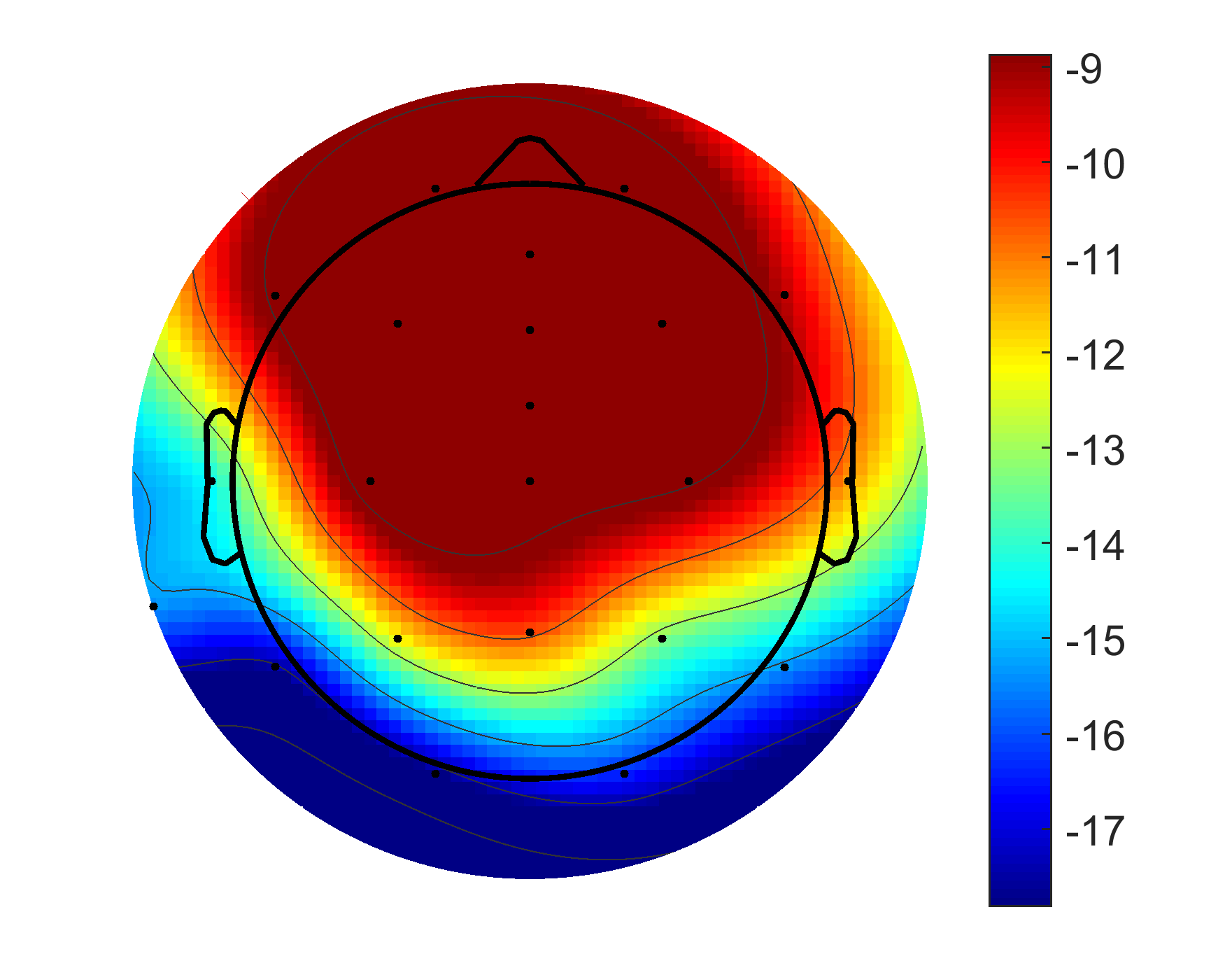}
\caption{\textsl{Heatplot depicting the SNR (in dB) of the AEPs obtained at different scalp locations when the reference electrode was placed at the right mastoid.}}
\label{fig:AEP_SNR}
\end{figure}

\paragraph{Spatial Noise Covariance Estimation} 
Since EEG amplifiers are differential amplifiers, they amplify the difference between signals measured at the data electrodes and the reference electrode. As a result, the differential component reduces as data electrodes are placed closer to the reference electrode. This can be better understood looking at auditory evoked potentials (AEPs) at different scalp locations \cite{bib:wolpaw1982}. AEPs are a convenient tool to analyze SNR characteristics as we know the true noiseless signal beforehand due to ensemble averaging. This is not the case in speech tracking as we are working on single trial and ensemble averaging is not feasible. The peak amplitude of AEPs around 100 ms is termed as N1 peak and it is a measure of the quality of the generated AEP \cite{bib:burkard2007}. Given that the reference electrode is placed at mastoid, N1 peak is largest in the vertex locations and reduces as we move towards the temporal locations in both directions. Fig. \ref{fig:AEP_SNR} depicts the SNR distribution of the AEPs obtained at different scalp locations. SNRs are usually between -9 dB to -17 dB and electrodes closer to the reference electrode have low SNR compared to the vertex electrodes \cite{bib:kuruvila2020} \cite{bib:hoppe2001}. Consequently, we can use signals at electrodes closest to the reference electrode to calculate the covariance matrix of noise that is required to solve \eqref{eq4} or \eqref{eq6}. This is under the assumption that signal to noise characteristics remain similar between AEP and AAD paradigm as the stimulus in both scenarios are acoustic signals.

If EEG signal ($w$) only from a single electrode is available for noise estimation, then the noise covariance matrix can be calculated as

\begin{equation} \label{eq8}
C_w = \sigma^2(w)I,
\end{equation}
where $\sigma^2(\cdot)$ represents the variance and $I$ is an identity matrix of order $N\times N$. In \eqref{eq8}, we assume that the noise is sampled from a set of uncorrelated random variables that are identically distributed.

If $L>1$ electrodes are available for noise estimation, then $C_w$ can be calculated as

\begin{equation} \label{eq9}
C_w = \frac{1}{L-1} \sum \limits_{n=1}^{L}(w_k - E(w_k))(w_k - E(w_k))^T.
\end{equation}
To ensure invertibility of the matrix in \eqref{eq4} and \eqref{eq6}, rank of $C_w$ must be at least $N$.\\

\paragraph{Noise Covariance Estimation in Multispeaker Scenario} 
When multiple speakers are present in an auditory scene, cortical signals measured at an EEG electrode can be influenced by every speaker present. In a dual-speaker scenario, the observed EEG signal $r$ can be written as

\begin{equation} \label{eq10}
r = \theta_a*s_a + \theta_u*s_u +w,
\end{equation}
where $\theta_a$ represents the attended TRF, $\theta_u$ represents the unattended TRF, $s_a$ represents the attended speaker envelope and $s_u$ represents the unattended speaker envelope. $\theta_a$ and $\theta_u$ can be estimated separately using \eqref{eq5}-\eqref{eq7}. However, in doing so, we assume the other speaker as noise in the EEG signal. That means, during the estimation of $\theta_a$, the cortical signal generated from the unattended speaker is considered as interference and is added to the noise and vice versa. Subsequently, for the estimation of $\theta_a$, \eqref{eq10} is rewritten as

\begin{equation} \label{eq11}
r = \theta_a*s_a + w_a
\end{equation}

where $w_a = w + \theta_u*s_u$,\\

and for the estimation of $\theta_u$, \eqref{eq10} is rewritten as

\begin{equation} \label{eq12}
r = \theta_u*s_u + w_u
\end{equation}

where $w_u = w + \theta_a*s_a$.\\

Finally, we make an approximation such that
\begin{equation} \label{eq13}
\sigma^2(w_a) \approx \sigma^2(w_u) = \sigma^2(w).
\end{equation}

\begin{figure}[h]
\centering
\includegraphics[width=.5\textwidth]{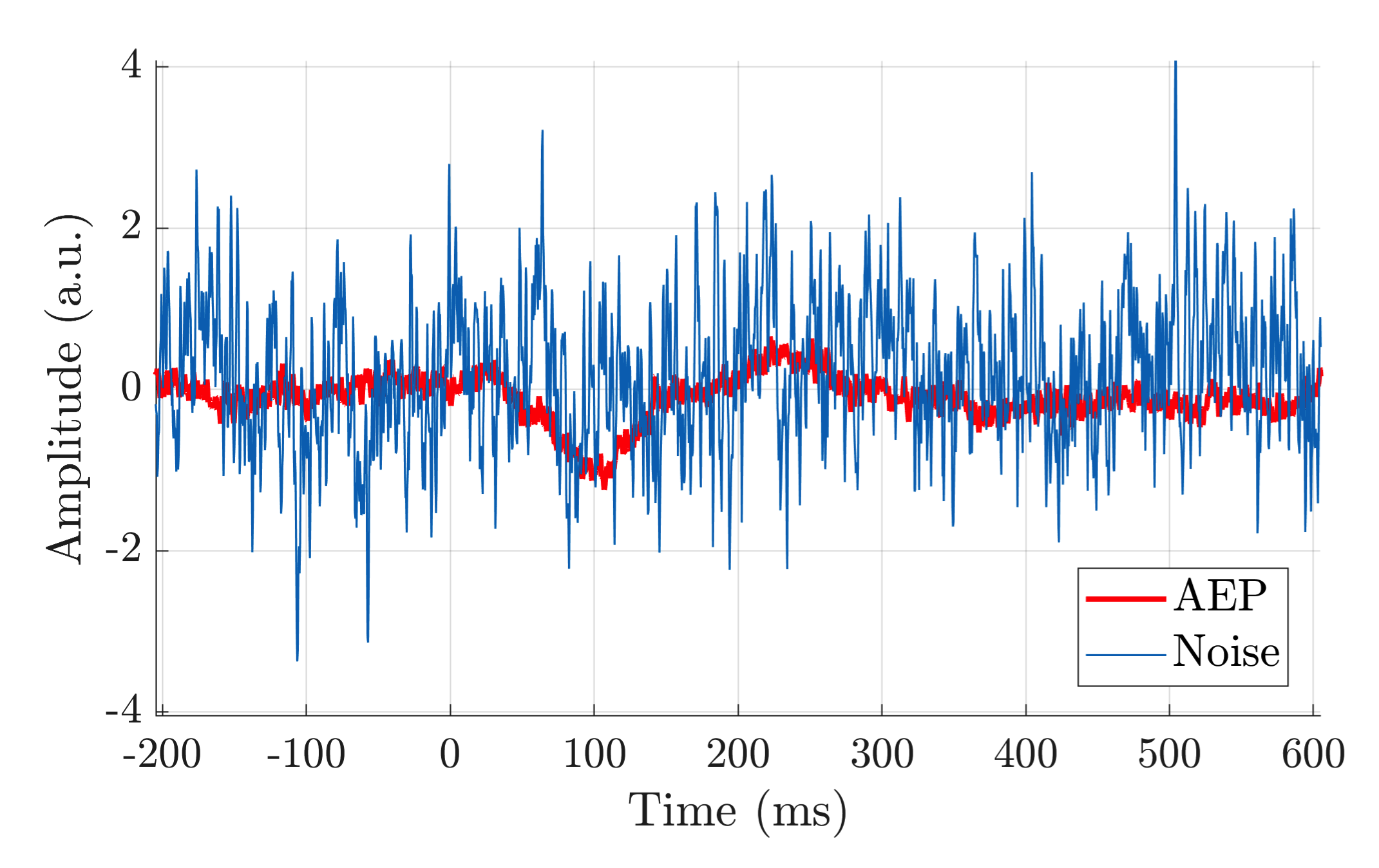}
\caption{\textsl{Plot showing the true signal (AEP) and the background noise (SNR = -10.8 dB) for a single epoch measured at Cz electrode from a representative subject. AEP was obtained by averaging over 100 epochs.}}
\label{fig:AEP_10_8dB}
\end{figure}

The rational for this approximation is as follows. The component of EEG that should have been contributed by the acoustic stimulus is significantly small compared to the background noise. Hence, the error introduced in the noise variance calculation by assuming interference from the other speaker as noise should also be significantly small. This is illustrated for the case of AEP in Fig. \ref{fig:AEP_10_8dB}.

\subsection{Extraction of Attention Markers}
\label{sec:EAM}
Attention marker is a measure of the reliability of estimated TRFs in decoding the attentional states. The correlation based attention marker is a commonly used measure which is calculated as the Pearson correlation coefficient between the original signal and the reconstructed signal. To be precise, once the TRF is estimated, it is used to reconstruct the EEG signal as a linear convolution between the speech envelope and estimated the TRF. As there should be at least two speakers present for selective attention, two separate EEG signals can be reconstructed using the speech envelopes and the estimated TRF. Then the speaker corresponding to the reconstructed EEG having a larger Pearson correlation coefficient to the original EEG is flagged as the attended speaker. Alternatively, correlation based attention marker can be calculated using backward model where the speech envelope is reconstructed from the EEG signals \cite{bib:sullivan2014}.

It has been shown that the correlation based attention marker is highly fluctuating for trial durations of the order of seconds resulting in an unreliable estimate of attention \cite{bib:miran2018} \cite{bib:geirnaert2019}. Hence, we decided to use the magnitude of N1\textsubscript{TRF}$-$P2\textsubscript{TRF} peak of the estimated TRFs as the attention markers. N1\textsubscript{TRF} is the negative peak that occurs around 100 ms latency and P2\textsubscript{TRF} is the positive peak that occurs around 200 ms latency and they are known to modulate attentional effect  \cite{bib:fiedler2019} \cite{bib:ding2012b} \cite{bib:akram2016a}. In our analysis, 75 ms - 135 ms latency was chosen as the time frame to search for N1\textsubscript{TRF} and 175 ms - 250 ms latency was chosen as the time frame to search for P2\textsubscript{TRF}. If there was no negative peak present in 75 ms - 135 ms time frame, N1\textsubscript{TRF} was initialized to zero. Similarly, if there was no positive peak present in 175 ms - 250 ms time frame, P2\textsubscript{TRF} was initialized to zero.

\subsection{Classification using SVM}
\label{sec:SVM}
For a linearly separable binary classification problem, SVMs aim to find a unique separating hyperplane that maximizes the margin between two classes. Searching the best separating hyperplane describes an optimization problem that can be solved efficiently using existing convex optimization algorithms. Furthermore, it can be extended to linearly non-separable classes by adding slack variables to the optimization problem as an additional constraint.        

Let the separating hyperplane that describes the decision boundary be defined as a first order affine function $f(\textbf{x})$ such that

\begin{equation} \label{eq14}
f(\textbf{x})=a^T\textbf{x}+b.
\end{equation}
It can be solved as \cite{bib:cortes1995}

\begin{equation} \label{eq15}
a =  \sum_{i=0}^m\alpha_i y^{(i)}\textbf{x}^{(i)},
\end{equation}
where $\alpha_i$ is a Lagrangian multiplier and $\textbf{x}^{(i)}$ and $y^{(i)}$ are the feature vector and the class label of the training samples respectively. Substituting \eqref{eq15} into \eqref{eq14}, the decision boundary can be rewritten as

\begin{equation} \label{eq16}
f(\textbf{x}) =  \sum_{i=0}^m\alpha_i y^{(i)} \langle \textbf{x}^{(i)},\textbf{x} \rangle + b.
\end{equation}

The Lagrangian multiplier $\alpha_i$ is non zero only for those $\textbf{x}^{(i)}$ and $y^{(i)}$ that act as support vectors \cite{bib:hastie2001}. The inner product in \eqref{eq16} is commonly referred to as kernel function and they help SVMs to map the input vector to higher dimensions. There are many choices for the kernel function such as linear, polynomial or Gaussian kernel. In our work, we used a linear kernel function to train the SVM. Fitting a sigmoid function after computing SVM output enables us to obtain posterior probabilities \cite{bib:platt1999}. In this work, SVM training was performed using \textit{fitcsvm} API and logistic regression training was performed using \textit{fitposterior} API provided in the MATLAB statistical and machine learning toolbox.
 
\subsection{Estimating Model Parameters}
\label{sec:EMP}
\begin{figure}[h]
\centering
\includegraphics[width=.5\textwidth]{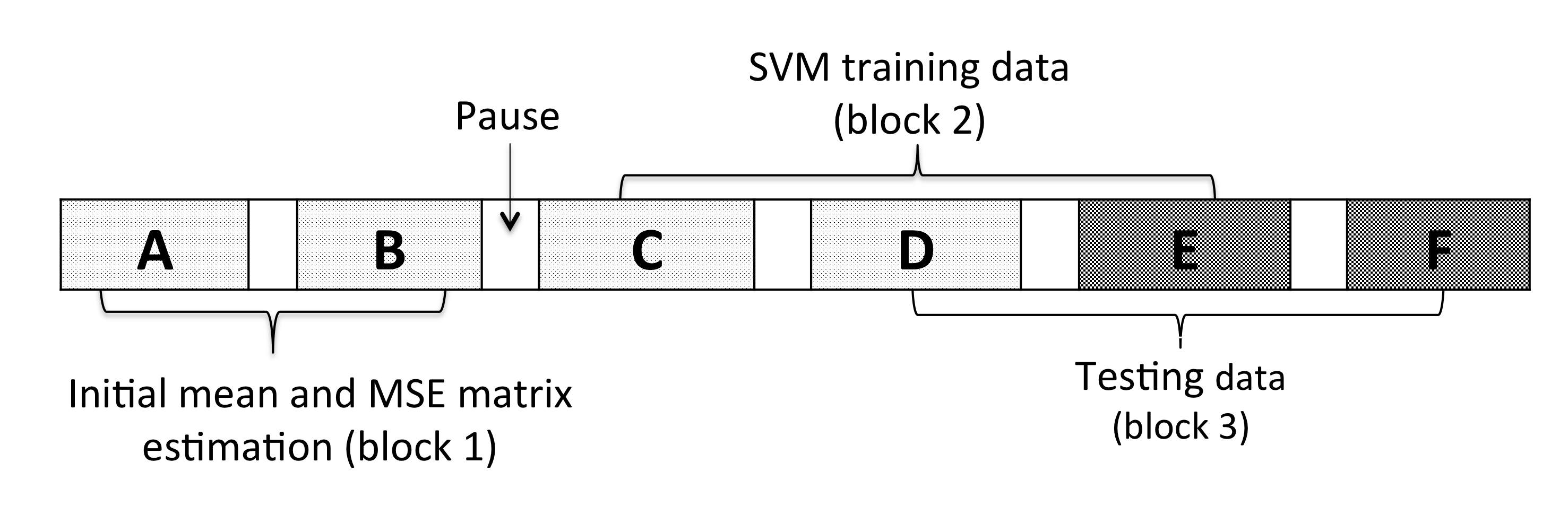}
\caption{\textsl{Data processing sequence. EEG was collected in six separate segments. In the first four segments, subjects maintained their attention to the same speaker. For the fifth and the sixth segments, they were asked to switch their attention to the other speaker. Segments A and B were used to calculate the initialization parameter, segments C and E were used for the SVM training and segments D and F were used to test the trained SVM classifier.}}
\label{fig:EMP}
\end{figure}

As described in section \ref{sec:ED}, our dual-speaker EEG experiment consisted of six segments (A-F), each being five minutes long. During the first four segments, subjects focused their attention to the same speaker and during the remaining two segments, subjects switched their attention to the other speaker. In order to calculate $\hat{\theta}[-1]$, $M[-1]$ which is needed to solve \eqref{eq5} - \eqref{eq7}, we split the EEG data into three blocks as shown in Fig. \ref{fig:EMP}. The first block consisted of segments A and B and it was used to calculate the initialization parameter $\hat{\theta}[-1]$ and $M[-1]$. The second block consisted of segments C and E and it was used to train the SVM classifier. Finally, the third block consisted of segments D and F and it was used to validate the performance of the trained SVM classifier. As the second and the third blocks comprised of attention to both speakers, we were able to test the ability of the algorithm to track attention switch.

In \cite{bib:kuruvila2020}, initialization parameters $\hat{\theta}[-1]$ and $M[-1]$ were chosen as zero mean and unit variance respectively. However, we decided to use the first block to calculate $\hat{\theta}[-1]$ and $M[-1]$. We started with the zero mean and the unit variance assumption and solved \eqref{eq5}-\eqref{eq7} sequentially for every two seconds trial. Once all the trials were completed, estimated $\hat{\theta}$ and $M$ in the last trial of the first block were used to initialize $\hat{\theta}[-1]$ and $M[-1]$ for the second block. It must be noted that, as there were two speakers present, we obtained two sets of $\hat{\theta}[-1]$ and $M[-1]$ in the last trial of first block. Hence, we took their average as the initial value. Alternatively, if EEG data corresponding to single-speaker attention is available, it could be used to perform this initialization. In the second block, as part of the SVM training, TRFs corresponding to the two speakers were estimated and their N1\textsubscript{TRF}$-$P2\textsubscript{TRF} peaks were extracted. The N1\textsubscript{TRF}$-$P2\textsubscript{TRF} peak served as the attention marker and it was labeled with the class labels corresponding to the attended and the unattended speaker. Finally, these labeled attention markers were used as feature vectors to compute the SVM decision boundary. In the third block, TRFs corresponding to the new data were estimated and their N1\textsubscript{TRF}$-$P2\textsubscript{TRF} peaks were passed as input to the trained SVM model to classify attention.

To compare the performance of our AAD framework with existing algorithms, attention decoding accuracies were calculated using LS method via stimulus reconstruction \cite{bib:sullivan2014} and state space model \cite{bib:miran2018}. In case of the LS method, the first and the second blocks were used to estimate the decoder and find the optimal ridge regularization parameter. The third block was used to infer the attention using the estimated decoder. During the decoder estimation, the autocorrelation matrices were averaged across trials to obtain a single final decoder \cite{bib:biesmans2017} instead of averaging the decoders estimated across multiple trials \cite{bib:sullivan2014}. We performed two variants of the LS method using stimulus reconstruction: 1) \textit{LS\_2sec} and 2) \textit{LS\_60sec\_overlap}. For \textit{LS\_2sec}, the trial duration considered to estimate the decoder was two seconds whereas for \textit{LS\_60sec\_overlap}, the trial duration considered was 60 seconds but with a sliding window of two seconds. I.e., in every trial, new data of only two seconds were added. 

In case of the state space model, first and second blocks were used to estimate the decoder and extract the attention markers for every two seconds trial. Once attention markers were extracted, they were used to fit a log-normal distribution and the mean and the variance of log-normal distribution were used to train the state space model. Finally, the third block was used to infer the attention using the trained state-space model. The algorithm proposes two different attention markers namely the Pearson correlation coefficients of the reconstructed stimuli and the L1 norm of the estimated decoders. In our comparative analysis, the L1 norm of the estimated decoder was used as the attention marker as it has been shown to have superior performance over the Pearson correlation coefficient in the original paper. All other hyperparameters were initialized to the default values. More details of the algorithm can be found in \cite{bib:miran2018}.

\section{Results}
\label{sec:Results}
\subsection{TRF Estimation : Sequential LMMSE vs LS}

\begin{figure}[b]
\centering
\includegraphics[width=.5\textwidth]{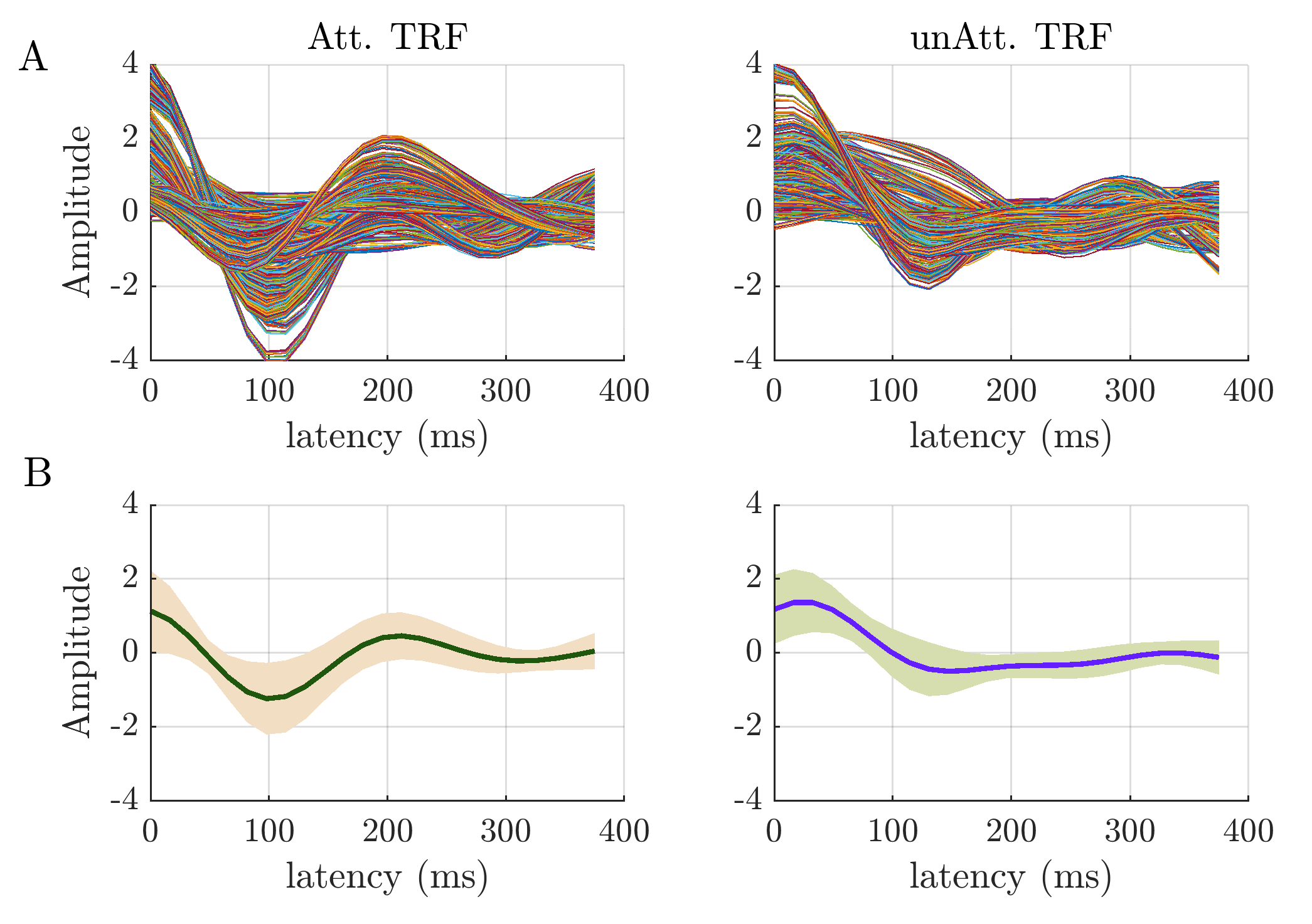}
\caption{\textsl{Sequential LMMSE estimate: Upper row: Attended TRF (left) and unattended TRF (right) estimated at Cz electrode from a representative subject. Each coloured line indicates the TRF obtained from separate trials that were of two seconds duration. Lower row: Mean $\pm$ std\_dev across trials.}}
\label{fig:sub_6_TRF}
\end{figure}

\begin{figure}[h]
\centering
\includegraphics[width=.5\textwidth]{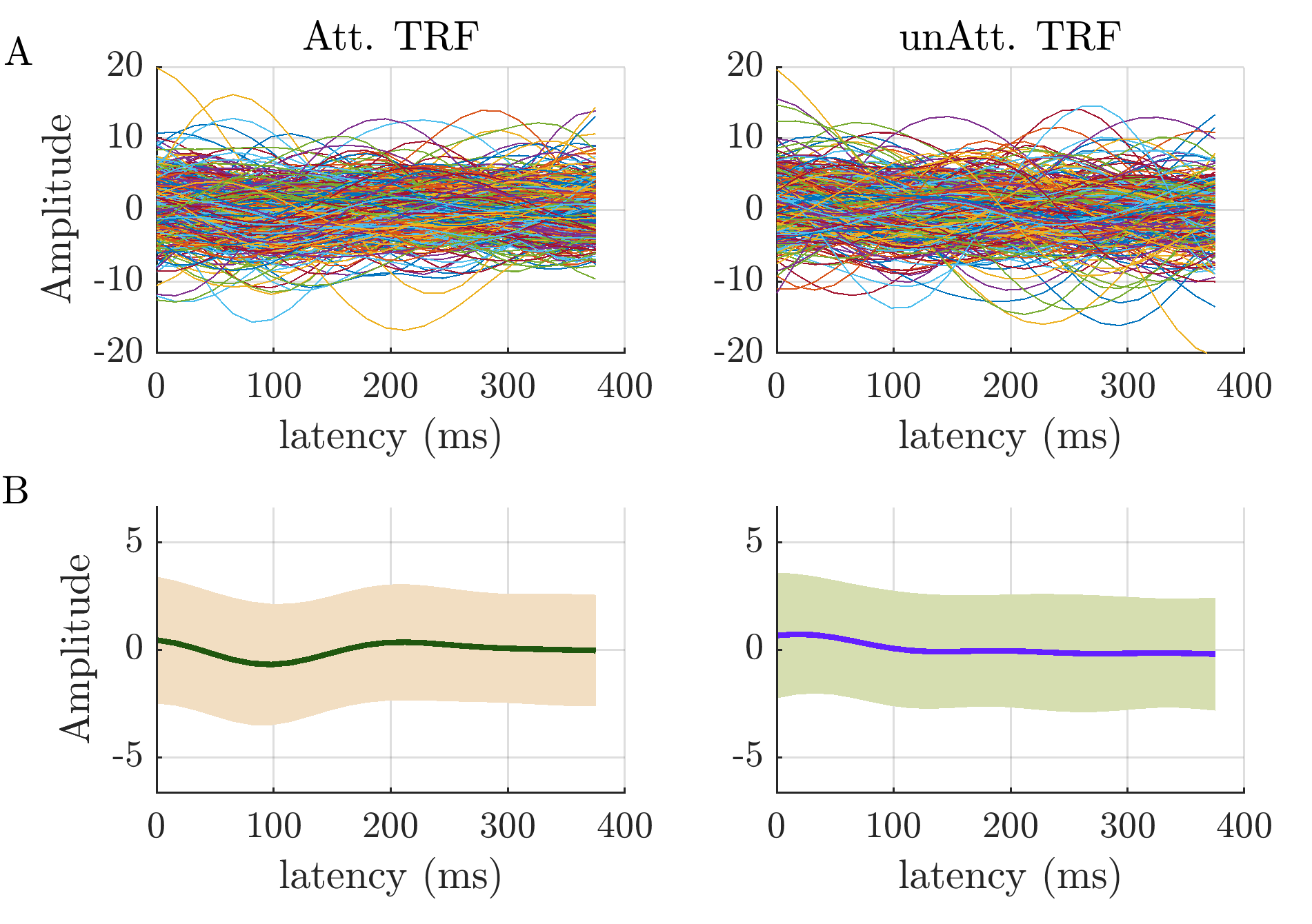}
\caption{\textsl{LS estimate: Upper row: Attended TRF (left) and unattended TRF (right) estimated at Cz electrode from the same subject as in Fig. \ref{fig:sub_6_TRF}. Each coloured line indicates TRF obtained from separate trials of two seconds duration. Lower row: Mean $\pm$ std\_dev across trials.}}
\label{fig:sub_6_TRF_LS}
\end{figure}

\begin{figure}[h]
\centering
\includegraphics[width=.45\textwidth]{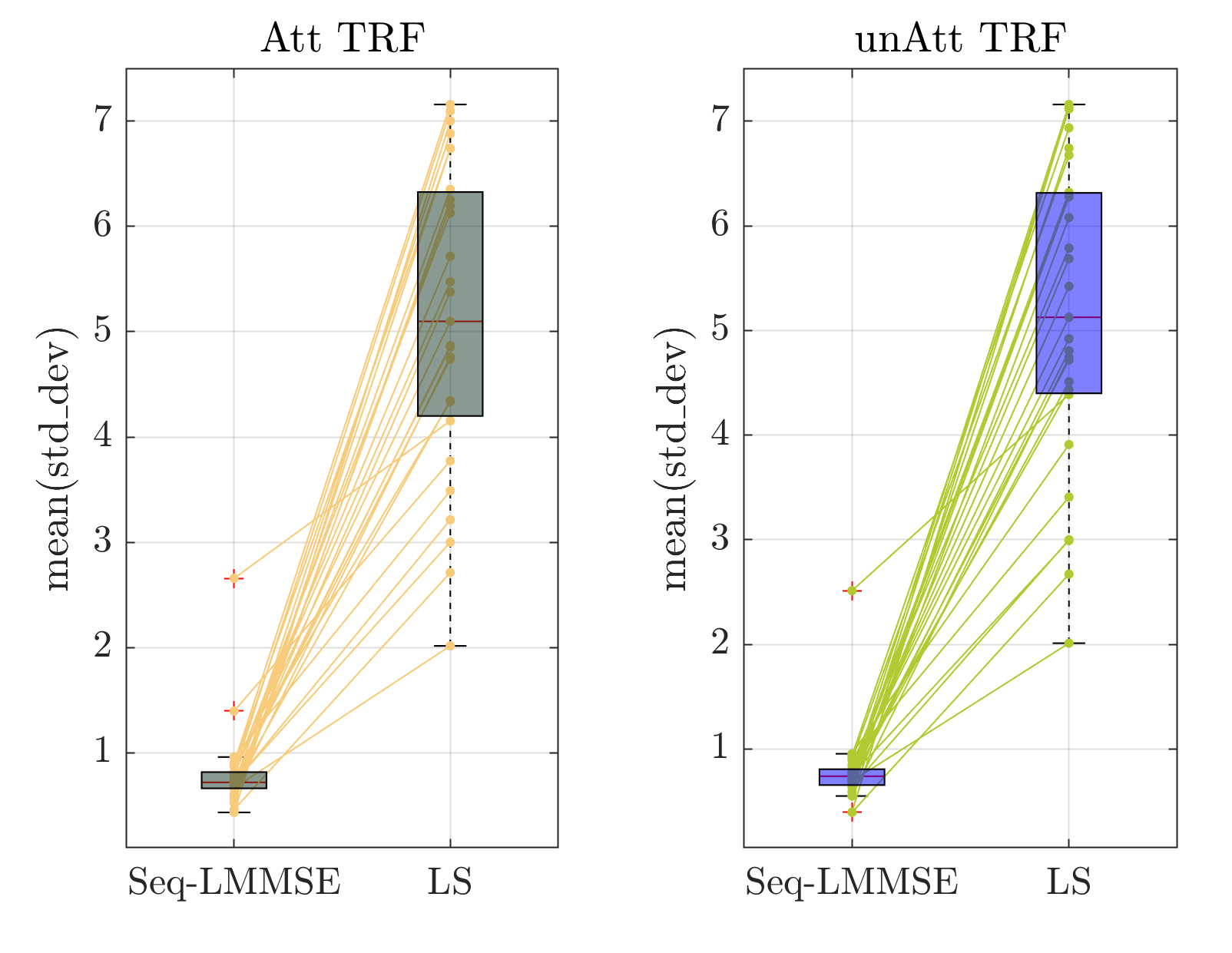}
\caption{\textsl{Comparison of the mean standard deviation of the attended and the unattended TRFs estimated using sequential LMMSE and LS algorithms. Each dot on the boxlot represents an individual subject.}}
\label{fig:est_std_dev}
\end{figure}

In Fig. \ref{fig:sub_6_TRF}, the attended TRF (left) and the unattended TRF (right) estimated using the sequential LMMSE algorithm for a particular subject are shown. TRFs were estimated in every trial and for a filter lag of 24 samples that corresponded to 375 ms at 64 Hz sampling rate. For the same subject, TRFs estimated using the LS algorithm are shown in Fig. \ref{fig:sub_6_TRF_LS}. It can be observed that for the case of LS algorithm, the TRF estimated across trials are noisy with high variance. For all subjects considered, the mean of standard deviations of the TRFs estimated using sequential LMMSE and LS algorithms are displayed in Fig. \ref{fig:est_std_dev}. The mean value was calculated by averaging the standard deviations across each individual time lags per subject. It is evident that the sequential LMMSE algorithm resulted in lower standard deviation compared to the LS algorithm ($p_{att}, p_{unatt} < 0.001$ based on paired Wilcoxon signed-rank test).  

%\begin{figure*}[t]
\begin{figure*}
\centering
\includegraphics[width=.78\textwidth]{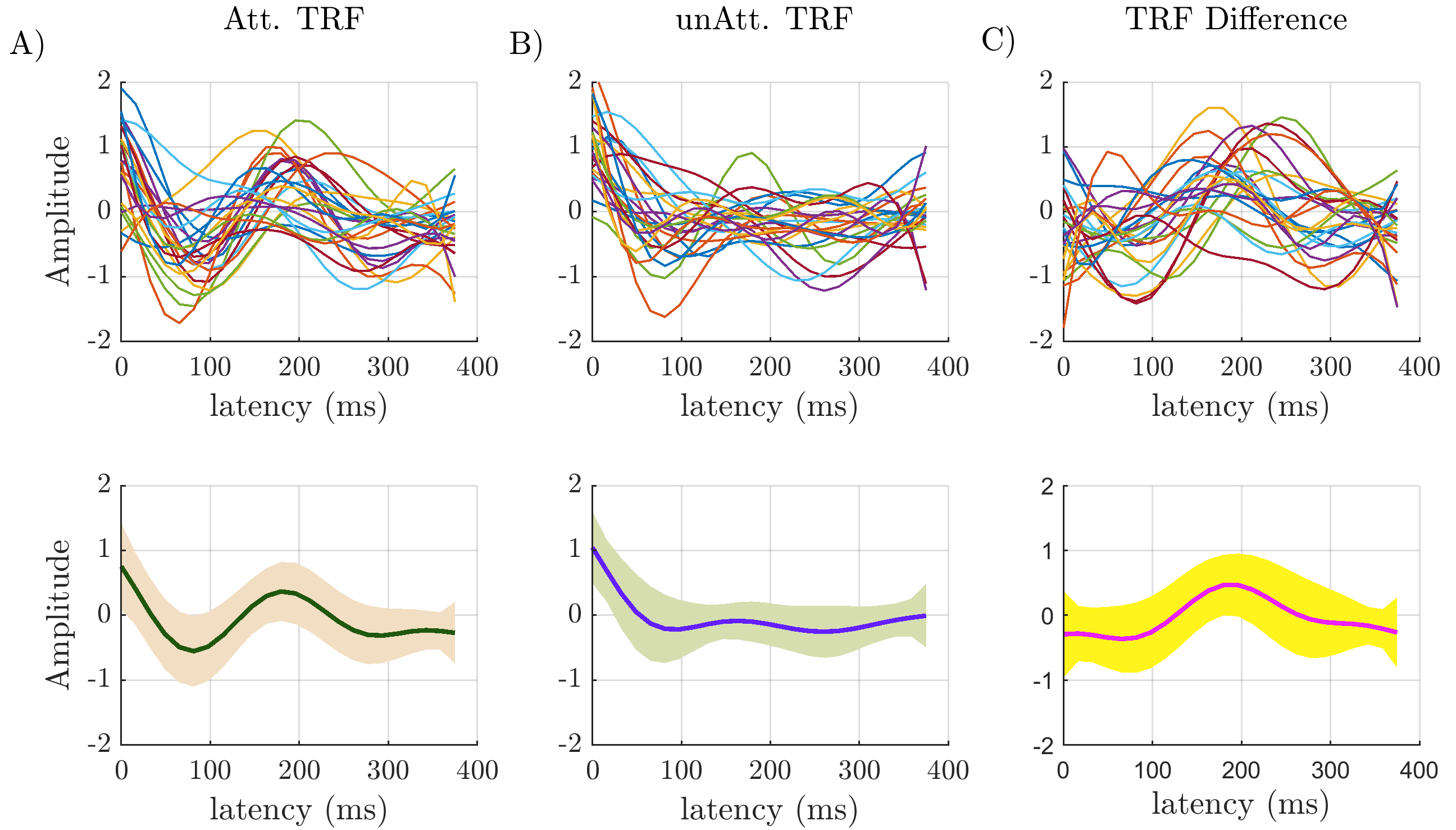}
\caption{\textsl{Plots depicting the attended TRF (A), the unattended TRF (B) and the TRF difference (C) estimated at Cz electrode for every subject. The upper row corresponds to the mean TRF of each individual subject and the lower row corresponds to the mean $\pm$ std\_dev across subjects. Each individual TRF on the upper row was obtained as an average of the TRFs across different trials.}}
\label{fig:att_unatt_TRFs}
\end{figure*}

%\begin{figure*}[t]
\begin{figure*}
\centering
\includegraphics[width=.94\textwidth]{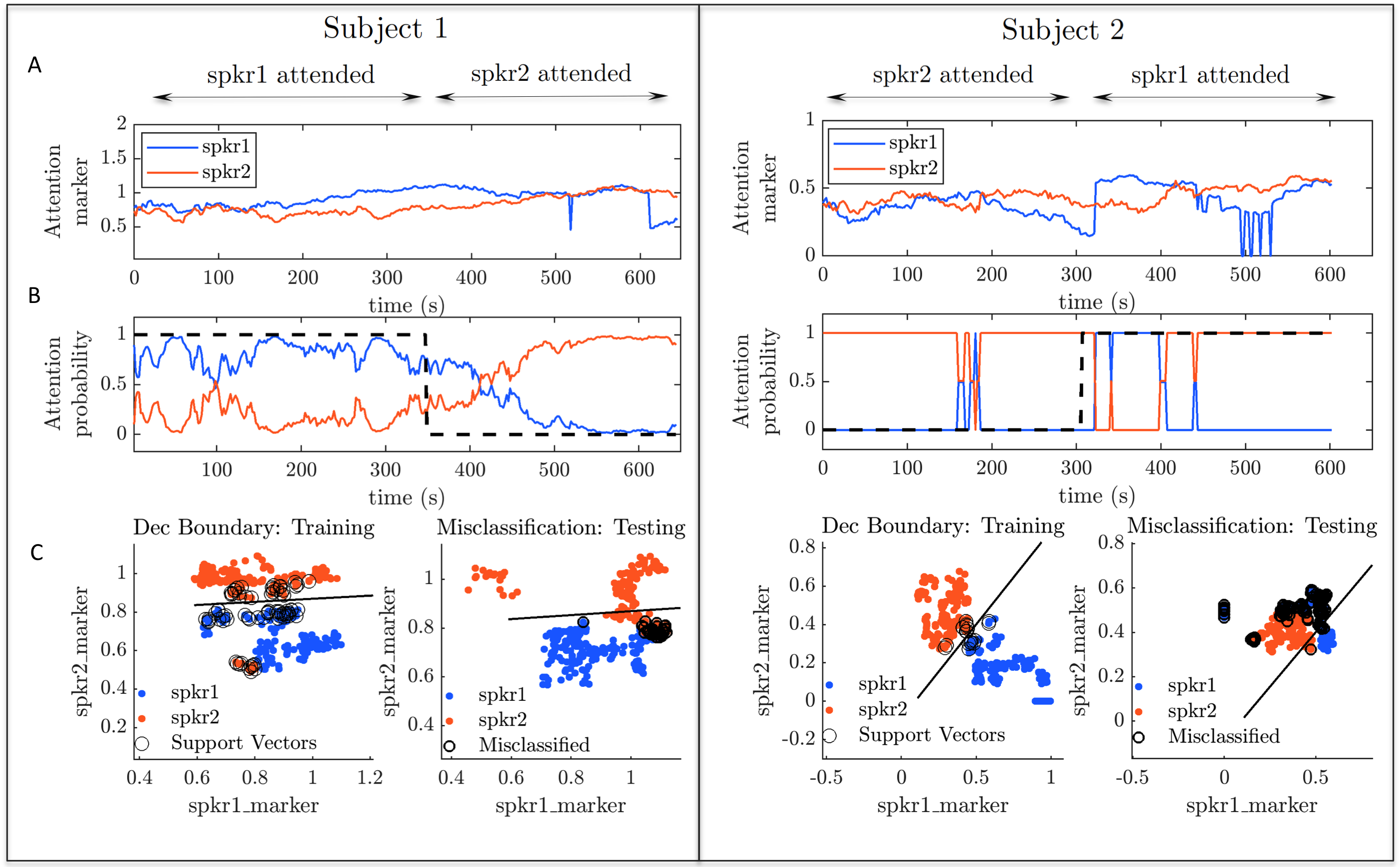}
\caption{\textsl{Decoding of the selective auditory attention of two representative subjects. For subject 1 (left panel), the initial attention was to speaker 1 and for subject 2 (right panel), the initial attention was to speaker 2. A) Dynamic evolution of the attention markers (N1\textsubscript{TRF}$-$P2\textsubscript{TRF} at Cz electrode) corresponding to the two speakers. A new attention marker was extracted for every trial of two seconds. B) Probability of attention to speaker 1 and speaker 2. Dotted line indicates the true attention to speaker 1. Attention switch was not a continuous process, instead it was obtained by concatenating two EEG segments with opposite attention (refer section \ref{sec:EMP}). The attention probability was calculated by passing the attention markers to an SVM followed by a logistic regression. The attention probability is dependent on the width of the SVM decision boundary that is determined by the spread of the support vectors. C) Left scatter plot depicts the linear decision boundary separating the attention markers. The abscissa (spkr1\_marker) corresponds to the N1\textsubscript{TRF}$-$P2\textsubscript{TRF} peak extracted from the TRF of speaker 1 and the ordinate (spkr2\_marker) corresponds to the N1\textsubscript{TRF}$-$P2\textsubscript{TRF} peak extracted from the TRF of speaker 2. Each shaded circular dot is a 2D vector with its first component being spkr1\_marker and its second component being spkr2\_marker. On average, when attending to speaker 1, the first component should be larger than the second component and vice versa. The unshaded black circles on the left scatter plot depict support vectors that define the decision boundary. The right scatter plot shows attention markers obtained during the testing procedure with black circles indicating misclassifications.}}
\label{fig:SVM_Cz}
\end{figure*}

\subsection{Attended vs Unattended TRF}
TRFs obtained from all the subjects with attended TRF (A), unattended TRF (B) and the difference in magnitude between them (C) are summarized in Fig. \ref{fig:att_unatt_TRFs}. On the upper panel, TRFs from the individual subjects are displayed and on the lower panel, their mean $\pm$ standard deviation are displayed. From Fig. \ref{fig:att_unatt_TRFs}A and Fig. \ref{fig:att_unatt_TRFs}B, it is evident that there is a negative peak at around 100 ms (N1\textsubscript{TRF}) and a positive peak at around 200 ms (P2\textsubscript{TRF}). However, the magnitude of N1\textsubscript{TRF} and P2\textsubscript{TRF} are higher for the attended TRF compared to the unattended TRF. The difference between the attended TRF and the unattended TRF (Fig. \ref{fig:att_unatt_TRFs}C) is negative for N1\textsubscript{TRF} ($p < 0.01$) and positive for P2\textsubscript{TRF} ($p < 0.001$). I.e., the deflections at 100 ms are more negative and the deflections at 200 ms are more positive for attended TRFs compared to unattended TRFs. Similarly, for latencies below 50 ms, both TRFs displayed a clear positive amplitude with the unattended TRF having a slightly larger amplitude than the attended TRF ($p < 0.05$). The statistical tests presented here were performed using one-sample t-test.

\begin{figure}[b]
%\begin{figure}[t]
\centering
\includegraphics[width=.5\textwidth]{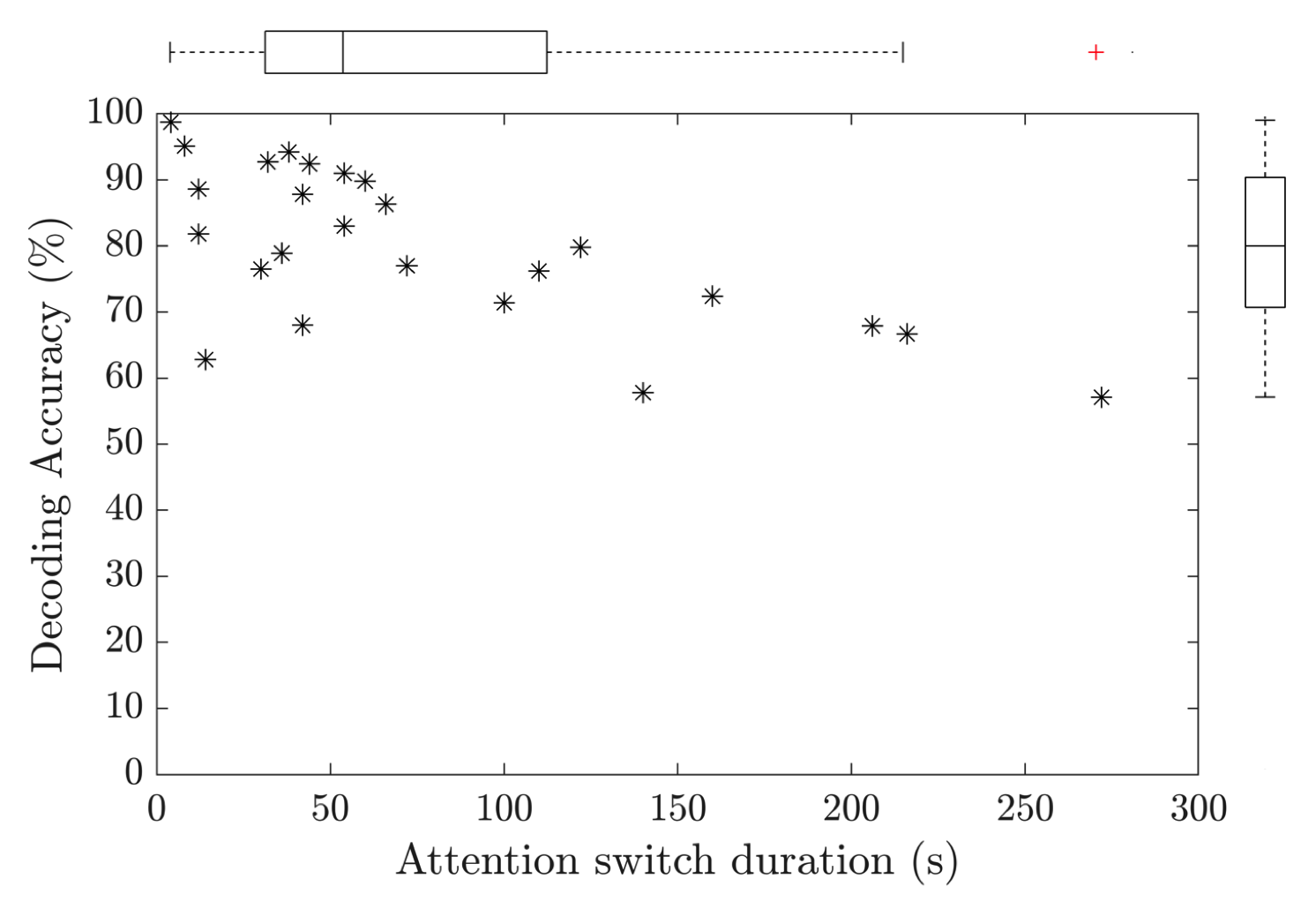}
\caption{\textsl{Decoding accuracy vs time taken to detect the attention switch for every subjects. Each * represents the result from an individual subject.}}
\label{fig:attn_switch_dur}
\end{figure}

\subsection{Decoding the Selective Attention}
Decoding the selective attention from two representative subjects is shown in Fig. \ref{fig:SVM_Cz}. For subject 1, the initial attention was to speaker 1 and the final attention was to speaker 2 whereas for subject 2, it was vice versa. For subject 1, the decoding accuracy was 89.5\% and for subject 2, the decoding accuracy was 62.8\%. The median decoding accuracy across all the subjects was 79.8\% which was above the statistical significance level. The statistical significance level for a trial duration of two seconds was found to be 54.7\% based on binomial distribution at 95\% confidence. Fig. \ref{fig:SVM_Cz}A displays the progression of attention markers (N1\textsubscript{TRF}$-$P2\textsubscript{TRF}) corresponding to the two speakers present. Fig. \ref{fig:SVM_Cz}B shows the probability of attention to speaker 1 which was obtained by passing the attention marker (Fig. \ref{fig:SVM_Cz}A) into a trained SVM classifier (Fig. \ref{fig:SVM_Cz}C). The probabilistic output gives a confidence in our attention inference and it is proportional to how far the attention markers are from the decision boundary. Regarding the multiple choice questionnaire, on average, subjects answered 84.7 $\pm$ 8.3\% of the questions correctly. Fig. \ref{fig:attn_switch_dur} depicts the latency that the algorithm required to detect the synthesized attention switch across all subjects. The attention switch duration ranged from as low as 4 seconds to as large as 272 seconds with a median value of 54 seconds. The attention switch duration also exhibited a negative linear correlation with the decoding accuracy ($r = -0.6765$).

\begin{figure}[t]
\centering
\includegraphics[width=.48\textwidth]{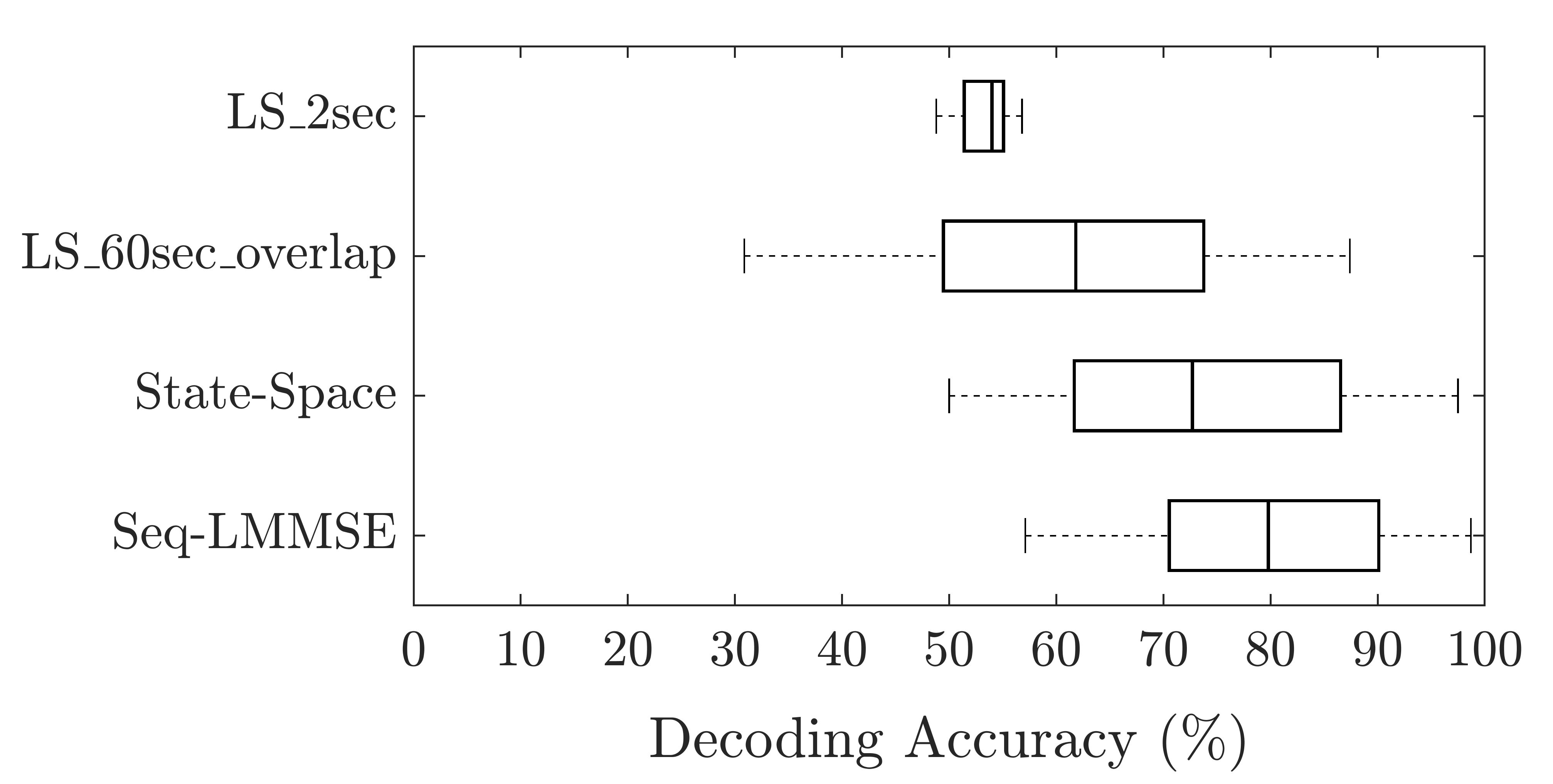}
\caption{\textsl{Comparison of the decoding accuracies obtained at Cz electrode using LS, state space and sequential LMMSE algorithms. The trial duration used was two seconds except for \textit{LS\_60sec\_overlap}.}}
\label{fig:comp_AAD_algo}
\end{figure}

\subsection{AAD Accuracy: Comparison with Existing Algorithms}
The median decoding accuracy achieved at Cz electrode for the case of \textit{LS\_2sec} was 53.96\% which was lower than the statistical significance level for a trial duration of two seconds (Fig. \ref{fig:comp_AAD_algo}). For \textit{LS\_60sec\_overlap}, median decoding accuracy at Cz electrode improved to 60.3\%. Here the trial duration considered was 60 seconds with a sliding window of two seconds. For the state space model, the median decoding accuracy using the same settings as \textit{LS\_2sec} was found to be 71.7\%. In addition, all aforementioned algorithms were tested using a combination of \textit{Cz + left mastoid} electrodes which was the same electrode combination used to estimate TRFs with sequential LMMSE. However, the accuracies were found to be slightly lower (not shown here) than the accuracies obtained using only Cz electrode.

\subsection{Spatial Distribution of TRFs}
\begin{figure*}[t]
%\begin{figure*}[h]
\centering
\includegraphics[width=1\textwidth]{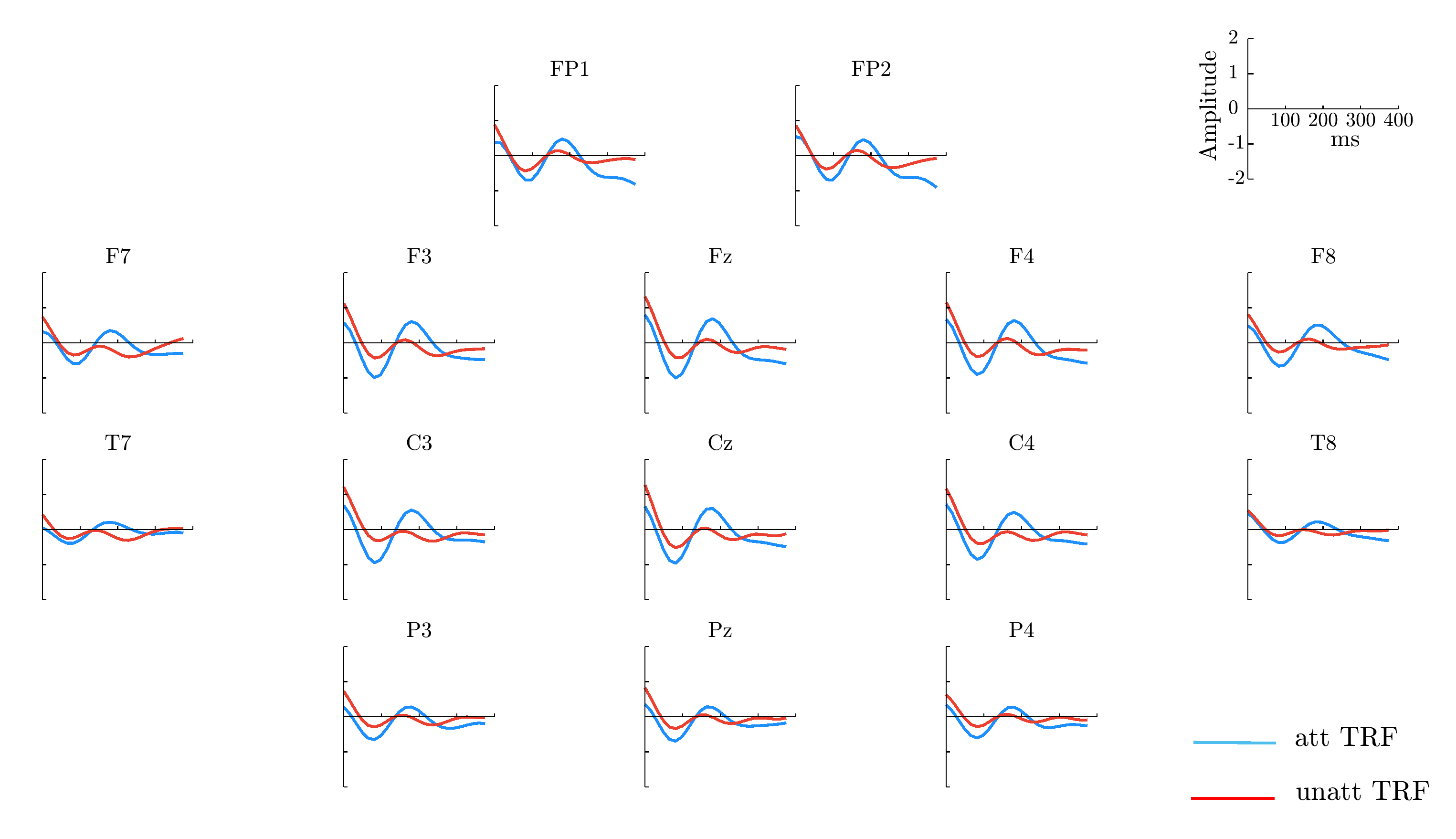}
\caption{\textsl{Spatial distribution of the attended TRF (blue) and the unattended TRF (red) estimated at different scalp locations. The TRF displayed at each location is the grand average obtained by calculating the mean across subjects. The reference electrode was placed at right mastoid.}}
\label{fig:all_electrodes}
\end{figure*}

Grand average TRFs estimated at different electrode locations referenced to the mastoid electrode are shown in Fig. \ref{fig:all_electrodes}. They were obtained by averaging the TRFs across individual subjects. As observed before, the amplitude of N1\textsubscript{TRF} and P2\textsubscript{TRF} corresponding to the attended TRF are larger than that of the unattended TRF. The magnitude of the N1\textsubscript{TRF}$-$P2\textsubscript{TRF} peak decreases as we move from the vertex regions to the temporal regions reminiscence of the SNR distribution of AEPs (Fig. \ref{fig:AEP_SNR}). The decoding accuracies obtained at different scalp locations are shown in Fig. \ref{fig:SVM_thresholding}. The highest decoding accuracy was obtained at Cz electrode (79.8 $\pm$ 12.68\%) and the lowest decoding accuracy was obtained at P4 electrode (66.2 $\pm$ 16\%).

\subsection{Accuracy Improvement with SVM}
\begin{figure}[h]
\centering
\includegraphics[width=.5\textwidth]{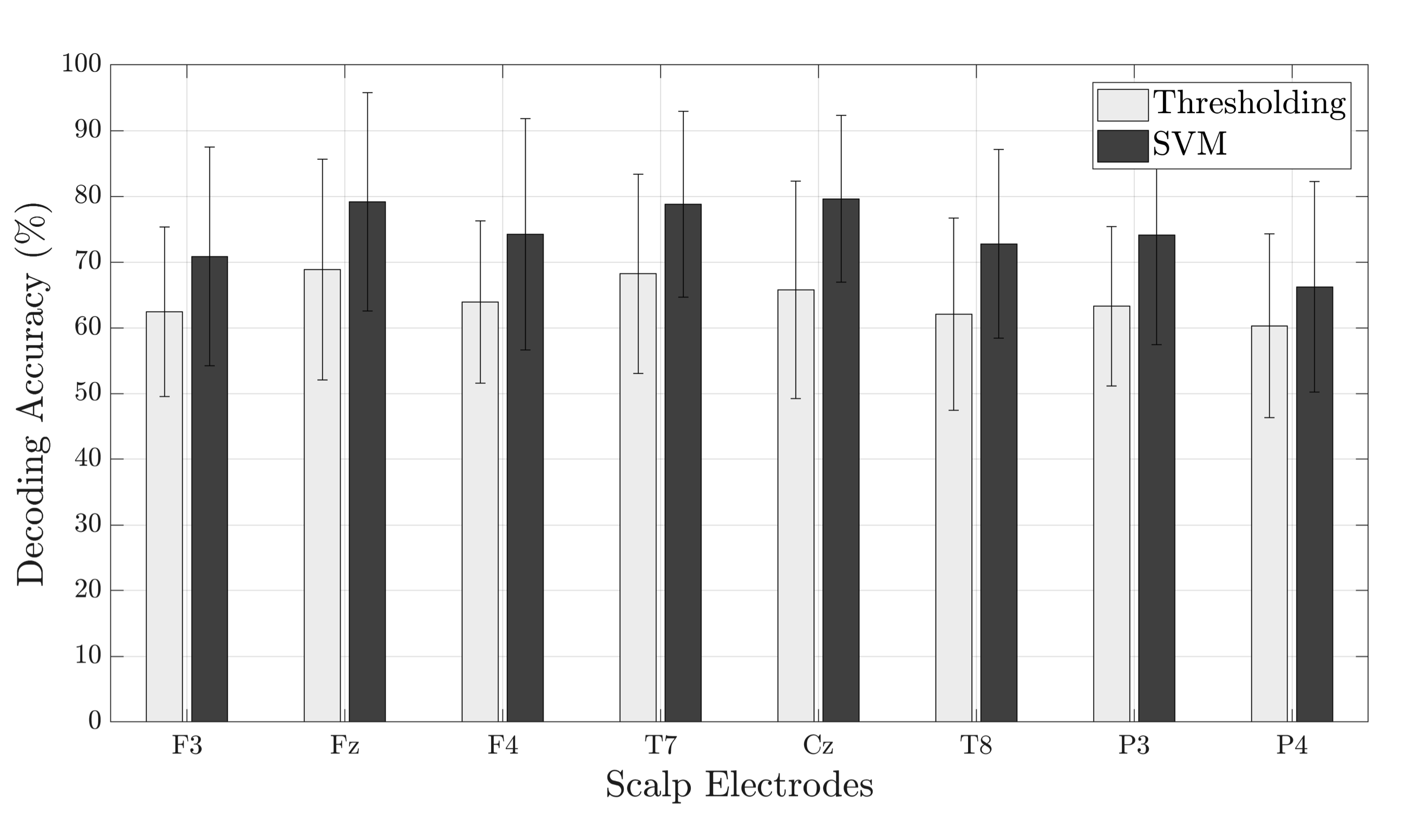}
\caption{\textsl{Comparison between the attention decoding accuracies obtained using SVM classifier and thresholding at selected scalp locations.}}
\label{fig:SVM_thresholding}
\end{figure}

SVM was used in our framework to classify the selective attention based on the N1\textsubscript{TRF}$-$P2\textsubscript{TRF} peak of estimated TRFs. To compare the improvements due to the introduction of SVM, decoding accuracies were calculated by thresholding the N1\textsubscript{TRF}$-$P2\textsubscript{TRF} peak. I.e., the speech envelope that generated the TRFs with larger N1\textsubscript{TRF}$-$P2\textsubscript{TRF} peaks was flagged as the attended speech. Fig. \ref{fig:SVM_thresholding} compares the decoding accuracies obtained using SVM and thresholding. At every electrode location evaluated, the accuracy was higher using the SVM classifier than thresholding. Mean of the accuracies obtained at different scalp locations using SVM was 74.5\% and that of using thresholding was 64.36\%.

\begin{figure}[h]
\centering
\includegraphics[width=.4\textwidth]{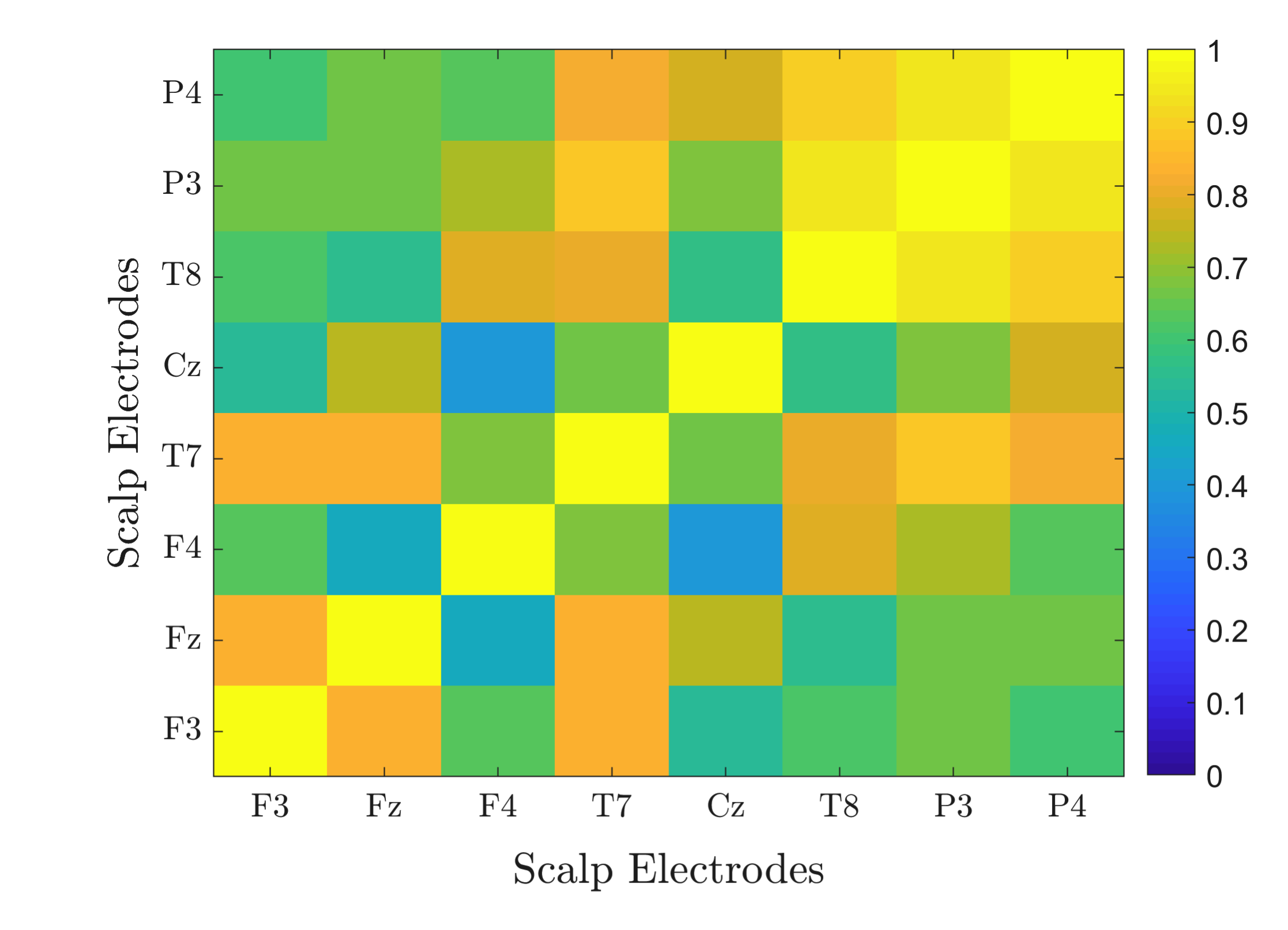}
\caption{\textsl{Mean correlation coefficient between the EEG signals (five minutes) measured at different electrode locations. The mean was obtained by averaging the correlation coefficients across twenty seven subjects.}}
\label{fig:sigCorr}
\end{figure}

\section{Discussion}
\label{sec:Discussion}
EEG analyses have been traditionally used to investigate the sensory response of human brain. In the auditory domain, AEPs which are responses evoked when stimulated by an acoustic signal are widely employed \cite{bib:burkard2007}. Due to the presence of substantial background noise, ensemble averaging must be performed to obtain AEPs that are visually interpretable. In other words, single trial analysis of EEG was a challenge until recently when it was demonstrated that the auditory attention in a dual-speaker scenario could be inferred without ensemble averaging using LS estimation \cite{bib:sullivan2014}. One of the limitations of the LS estimation was that the trial duration required to infer attention was of the order of minutes but a listener could switch the attention multiple time within time scales of this order. Several improvements were proposed by which the trial duration needed to infer attention was considerably reduced \cite{bib:miran2018} \cite{bib:deTaillez2017} \cite{bib:etard2019}.

In this study, we have proposed a framework to decode the selective auditory attention of a listener. The efficacy of the algorithm was verified using EEG data collected from 27 subjects who undertook selective auditory attention experiment. The proposed framework consists of three modules. In the first module, a novel TRF estimation algorithm is presented which is based on LMMSE estimator. The MMSE estimators are known to perform better than the LS estimators at low SNR and their performance converges as the SNR improves. The improvement stems from the fact that the MMSE estimators make use of second order statistics (covariance) of the noise present in the signal. To calculate the noise covariance matrix, we can use signals measured at locations closest or contralateral to the reference electrode. Another potential way of estimating the noise at a particular electrode is to measure the signal at that electrode before the stimulus is presented \cite{bib:das2020}. This is appropriate for experiments with non-continuous speech such as speech-in-noise test where the stimuli are presented with periodic pauses. When entire signals are not available a priori, sequential LMMSE estimator can be used to estimate the TRF. In sequential estimation, we assume that the TRF in the previous trial has an influence on the TRF in the current trial. The contribution of the previous estimate on the current estimate is controlled by the gain factor.

In the second module, a marker related to the attention of the listener is extracted. Amplitude peaks of the TRF around 100 ms (N1\textsubscript{TRF}) and 200 ms (P2\textsubscript{TRF}) are known to modulate selective attention \cite{bib:fiedler2019} \cite{bib:ding2012b} \cite{bib:akram2016a}. Hence, we used the N1\textsubscript{TRF}$-$P2\textsubscript{TRF} peak as the attention marker in our framework. Once the markers were extracted, they were passed to an SVM classifier followed by a logistic regression to obtain a probabilistic measure of the attention. Instead of using an SVM, the attention markers could be directly passed to the logistic regression but SVMs have the inherent ability to correct for errors if the feature vector falls inside the decision boundary. Alternate to SVMs, random forests or neural networks based classifiers could be used provided there is sufficient data available to train the network.  

\subsection{TRF Estimation : Sequential LMMSE vs LS}
Two commonly used linear system identification methods in statistical estimation theory are LS and LMMSE estimators. In LS estimator, the squared error is minimized without having any assumptions on the stochastic property whereas in LMMSE estimator, the mean squared error is minimized assuming a Gaussian PDF and taking the expectation over the assumed PDF. Algorithms based on LMMSE exploit the knowledge of noise covariance and in this work, we reformulated the covariance matrix of received EEG signal as a combination of the covariance matrix of the speech envelope and the covariance matrix of the noise signal. I.e., we used some prior knowledge about the covariance matrix and if this prior knowledge can be estimated with a sufficient accuracy, LMMSE algorithm outperforms LS algorithm \cite{bib:kay1997}. This performance improvement can be observed when comparing Fig. \ref{fig:sub_6_TRF} and Fig. \ref{fig:sub_6_TRF_LS}. The standard deviation (and variance) of the TRFs estimated using the LS estimator was found to be significantly larger than that of using the sequential LMMSE estimator (Fig. \ref{fig:est_std_dev}). However, the performance of the LS estimator should improve with increasing trial duration as it is known that the SNR of the LS estimate is directly proportional to the length of the trial \cite{bib:Scharf1991}. As a result, it is only when the length of the observation vector (EEG) significantly exceeds the length of the estimation vector (TRF), the LS estimate is reliable.

\subsection{Attended vs Unattended TRF}
Analysis of the filter coefficients of the attended and unattended TRFs suggest a top down processing of the auditory information. To be precise, a larger N1\textsubscript{TRF} and P2\textsubscript{TRF} for the TRF estimated using the attended speech envelope suggests that the brain might be suppressing the contents of the unattended speech from being encoded in the working memory. These results suggest that the LMMSE based TRF estimation could replicate the results obtained in previous studies \cite{bib:fiedler2019}  \cite{bib:ding2012b}. Responses beyond 100 ms are attributed to phoneme processing and suppression of the unattended TRF at these latencies means that the high level features corresponding to the unattended speaker are not encoded \cite{bib:diLiberto2015} \cite{bib:brodbeck2018}. For initial latencies below 50 ms, a large positive amplitude was observed for both attended and unattended TRFs. This is intriguing because previous studies have not reported large amplitudes at these latencies. A potential reason is the short trial duration that was used to estimate the TRF. I.e., as the stimuli in our experiment were continuous news segments that lasted around five minutes and as we were analyzing short trials of two seconds, subjects had an expectation about the incoming acoustic stimuli. These expectations may be getting reflected as early responses that are largely identical for both the attended and the unattended speaker. We anticipate that these early responses should diminish for non-continuous stimulus such as the one used in speech-in-noise tests.

In our study, analyses were limited to dual-speaker scenario but in real life, there could be more interfering speakers present. It has already been shown that the selective attention could be inferred in a four-speaker environment using the LS method \cite{bib:schafer2018}. TRF estimation using sequential LMMSE can be extended to multispeaker scenarios such as the four-speaker experiment to evaluate the behaviour of TRFs corresponding to each of the interfering speakers.

\subsection{Attention Decoding Accuracy}
The most important improvement introduced by the sequential LMMSE algorithm is its ability to generate interpretable TRFs from short duration trials. This has enabled us to move away from the conventional correlation based attention marker to the N1\textsubscript{TRF}$-$P2\textsubscript{TRF} peak based attention marker. Correlation based attention marker falls under the category of regression tasks because we have to reconstruct the EEG signals initially to calculate the Pearson correlation coefficient. However, inferring attention directly based on the N1\textsubscript{TRF}$-$P2\textsubscript{TRF} peak falls under the category of classification tasks. It is known that regression tasks are more challenging than classification tasks and we expect an improvement in performance using N1\textsubscript{TRF}$-$P2\textsubscript{TRF} peak instead of Pearson correlation coefficient.

The decoding accuracy obtained at Cz electrode for a trial duration of two seconds using LS estimator was found to be 53.96\%. An analysis of overlapping trials with a duration of 60 seconds but using a sliding window of two seconds improved the accuracy to 60.3\%. On the other hand, the decoding accuracy obtained using our AAD framework was 79.8\% (Fig. \ref{fig:comp_AAD_algo}). The lower decoding accuracy using LS estimator can be attributed to its dependency on the number of electrodes and the trial duration. I.e., as the trial duration and the number of electrodes reduce, decoding accuracy deteriorates \cite{bib:mirkovic2015}\cite{bib:fuglsang2017} \cite{bib:narayanan2019}. An attention decoding framework based on the state space algorithm reported comparable accuracies to that of our framework. However, adaptation of the same algorithm on our dataset yielded an accuracy of 72.7\% which is lower than the accuracy reported in \cite{bib:miran2018}. The lower accuracy may be due to the choice of hyperparameters in our adaptation because we optimized only the LASSO regularization parameter and the parameters of the log-normal distribution. All other hyperparameters were initialized with the default values suggested in \cite{bib:miran2018}. The trial duration was chosen as two seconds because previous studies have shown that the human brain can hold upto two seconds of independent auditory information in the working memory \cite{bib:baddeley1975}. In future work, the trial duration could be chosen as the listener's memory span and it can be estimated with the help of digit span tests.

Despite our decoding framework being designed to detect the attention at a time resolution of two seconds, the median attention switch duration was found to be 52.4 seconds which was larger than our expectation. There are two possible explanations for the delayed detection of attention switch. First, our experiment was designed in such a way that there was no dynamic attention switch present. Instead, the attention switch was synthesized by concatenating two EEG segments with opposite attention. Hence, it is possible that the subject required certain amount of time to focus the attention on a particular speaker at the start of the experiment. Second, due to the sequential nature of estimating the current TRF based on previous estimates, the algorithm may have introduced a delay by itself which is reflected in the longer switch duration. To disentangle the aforementioned possibilities, the algorithm must be further evaluated in scenarios where the subject has the flexibility to switch attention dynamically \cite{bib:presacco2019b}. %Furthermore, Kalman filtering based TRF estimation can be explored by which we can control the influence of previous estimates on the current estimate \cite{bib:kay1997}.

\subsection{Spatial Distribution of TRF}
Comparison of the TRFs obtained at different scalp locations shows that frontal and central locations generate identical TRFs. This is not trivial considering the fact that EEG signals measured at these locations are highly correlated (Fig. \ref{fig:sigCorr}). For our choice of the mastoid reference location, the magnitude of N1\textsubscript{TRF}$-$P2\textsubscript{TRF} peaks are largest at the central and the frontal regions and smallest for the temporal regions. On the contrary, if the reference electrode was chosen as Cz electrode, the polarity of TRF will get inverted and temporal regions will have the largest N1\textsubscript{TRF}$-$P2\textsubscript{TRF} peak. Despite the smaller peaks, a clear difference between the attended and the unattended TRFs could be observed at all locations. Decoding accuracies obtained at the temporal electrodes were not significantly lower than the decoding accuracies obtained at the central electrodes. This could pave the way to potentially integrate these algorithms in a neuro-steered hearing aid. Furthermore, these results suggest that we may not require high-density EEG measurements with multiple electrodes in paradigms such as the selective auditory attention.

One of the assumptions that we have made throughout this paper is the availability of clean speech envelope to estimate the TRFs. In practice, only noisy mixtures are available and speech sources must be separated before the envelope can be extracted. This is an active research field and algorithms are already available based on classical signal processing such as beamforming or based on deep neural networks \cite{bib:wang2018}. Another limitation of the sequential LMMSE algorithm in its current form is that it can be used to estimate the auditory system response only in the forward direction (speech to EEG). This is because the noise covariance matrix that we have made use of is available only in the forward direction. 

\section{Conclusion}
In this paper, we have proposed a method to estimate the TRF from EEG signals using sequential LMMSE estimator. Unlike the LS estimator, sequential LMMSE estimator is capable of generating reliable and interpretable TRFs from short duration trials. Analysis of the properties of the TRFs in dual-speaker scenario has revealed a locus of attention around 100 ms (N1\textsubscript{TRF}) and 200 ms (P2\textsubscript{TRF}). I.e., the TRFs corresponding to the attended speaker have a larger N1\textsubscript{TRF}$-$P2\textsubscript{TRF} peak than the TRFs corresponding to the unattended speaker. Using sequential LMMSE as the major building block, we developed a novel AAD framework to detect the auditory attention of a listener at a time resolution of two seconds. %Our AAD framework consisted of three modules. In the first module, attended and unattended TRFs are estimated using the sequential LMMSE estimator. In the second module, the N1\textsubscript{TRF}$-$P2\textsubscript{TRF} peaks are computed from the estimated TRFs that serve as the attention markers. In the third and final module, the attention markers are given to an SVM classifier followed by a logistic regression to obtain a probabilistic measure of attention.%
In the proposed AAD framework, only two electrodes (data measurement and noise measurement) in addition to the reference and the ground electrodes are sufficient to achieve a reliable decoding accuracy. Comparison of the results obtained at different scalp electrodes revealed that the AAD accuracies are in the similar range across different electrodes.

Although the AAD framework was designed to detect the attention from EEG trials of the order of seconds, it was observed that the time taken to detect the attention switch was longer than a single trial duration. Hence, further investigation is required to understand the fundamental reason behind it and reduce the attention switch duration.

\section*{Acknowledgment}
We convey our gratitude to all participants who took part in the study and would like to thank the student Laura Rupprecht who helped us with data acquisition.

{
\small
\addcontentsline{toc}{chapter}{Bibliography}
\bibliography{literature}
\bibliographystyle{IEEEtran}
}

\end{document}